\documentclass[15pt a4paper]{article}

\usepackage{latexsym}
\usepackage{amsmath,amssymb,amsfonts,amsthm}

\usepackage{graphicx}

\graphicspath{{./}{./figs/}}

\begin{document}

\title{Synchronization in weighted scale free networks \\ with degree-degree correlation}

\author{F. Sorrentino $^{1}${\footnote{\small Email: {\emph fsorrent@unina.it}}}, M. di Bernardo $^1$, G. Huerta Cu\'ellar$^2$ and  S. Boccaletti $^2$ \\
{\small $^1$Department of Systems and Computer Science, University
of Naples Federico II, Via Claudio 21, 80125 Naples, Italy}\\
{\small $^2$CNR-Istituto dei Sistemi Complessi, Via Madonna del
Piano, 10, 50019 Sesto Fiorentino, Florence, Italy}\\
 }

 \maketitle
\begin{abstract}
 \medskip
We study synchronization phenomena in scale-free networks of
asymmetrically coupled dynamical systems featuring degree-degree
correlation in the topology of the connection wiring. We show that
when the interaction is dominant from the high-degree to the
low-degree (from the low-degree to the high degree) nodes,
synchronization is enhanced for assortative (disassortative)
degree-degree correlations.

\end{abstract}

Recent years have witnessed increasing interest in the study of the
structure of complex networks. Networks are present ubiquitously in
the real world, including biological systems, genetic chains, social
relationships and artificial and engineering architectures
\cite{report}. A main quantity that characterizes the structure and
dynamics of those networks is the so called degree distribution
$P(k)$, the probability that a randomly chosen node within the
network has degree $k$, i.e. there are $k$ connections incident in
it. In most cases, this distribution has been found to obey a power
law $P(k) \sim k^{- \gamma}$, with $\gamma$ usually ranging between
2 and 3. The slow decay of the power-law distribution, when the
exponent is within the observed values, is responsible for the
presence of few high-degree nodes, usually termed as \emph{hubs},
which play a leading role in characterizing the network dynamical
behavior. Thus the network
structure (termed in this case as {\it scale free}) displays %has
high heterogeneity, with the presence of a few high-degree vertices
and many low-degree ones.

Beyond the specific interest in the study of the main topological
properties of such networks, there is an ongoing research effort to
understand how these statistical features affect the dynamical
processes taking place over them, as e.g.  epidemic spreading
\cite{Pa:Ve00a,Bo:Pa02,New02Epid}, communication models
\cite{Oh:Sa,korea,Gu:Ar02,Ar:dB:So} and spin systems
\cite{Do:Go:Me,Do:Go:Me2,Mo:New00a}. Among these, a relevant
phenomenon is the emergence of collective synchronized motion. This
corresponds to the case of several dynamical systems, starting from
different initial conditions appropriately coupled with a network
topology, stabilizing in a common evolution.

In this framework, we consider the case of networks of coupled
identical systems, that can be conveniently described by the
following set of ordinary differential equations:
\begin{equation}
\frac{dx_i}{dt}=f(x_i)-\sigma \sum_{j=1}^{N} {\mathcal L}_{ij}
h(x_j), \qquad i=1,2,...N, \label{eq:net}
\end{equation}
where ${x}_i \in {\Bbb R}^m$ is the $m$-dimensional vector
describing the state of the $i^{\text {th}}$ node, ${f}({x_i}) :
{\Bbb R}^m \to {\Bbb R}^m$ governs the dynamics of the $i^{\text
{th}}$ node, ${h}({x}) : {\Bbb R}^m \to {\Bbb R}^m $ is a vectorial
output function, $\sigma$ is a parameter ruling the coupling
strength, and $N$ is the number of nodes of the network. The
topological information on the graph is contained in the connection
matrix $\mathcal L$. Specifically $\mathcal{L}=\{\mathcal{L}_{ij}\}$
is defined as follows:
\begin{eqnarray}
\mathcal{L}_{ij}=-\mathcal{A}_{ij}\mathcal{W}_{ij}, \quad i \neq j
\nonumber\\
\mathcal{L}_{ij}=\sum_{j}\mathcal{A}_{ij}\mathcal{W}_{ij}, \quad
i=j,\label{L}
\end{eqnarray}

 where $\mathcal{A}$ is the irreducible
adjacency matrix associated to the network topology, that is
$\mathcal{A}_{ij}=\mathcal{A}_{ji}=1$, if there is a connection
between $i$ and $j$, $0$ otherwise. $\mathcal{W}$ is a positive
definite matrix containing the information about the weights over
the network connections, with $\mathcal{W}_{ij}$ giving a measure of
the strength of the interaction from node $i$ to node $j$.

Note that (\ref{eq:net}) and (\ref{L}) encompass a general and large
class of networks, allowing for non-linearity in the functions $f$
and $h$ and leaving complete freedom for the choice of the weights
over the network connections. The only constraint is represented by
the condition $\mathcal{L}_{ii}= -\sum_{j \neq i} \mathcal
{L}_{ij}$, which is necessary in order to guarantee that the
connection matrix $\mathcal{L}$ is zero row-sum.

When the connection matrix $\mathcal L$ is symmetric, the network
propensity to synchronization (the so-called network
synchronizability) can be studied rigorously by means of the Master
Stability Function (MSF) approach \cite{Pe:Ca}. Namely, the MSF
assesses the necessary conditions for the transverse stability of
the synchronization manifold $x_1=x_2=...=x_N=x_s$, whose existence
as an invariant set for the system in (\ref{eq:net}) is warranted by
the zero row-sum property of $\mathcal{L}$.

Indeed, let $\delta {x}_{i}(t)={x}_{i}(t)-{x}_s (t)$ be the
deviation of the $i^{\text {th}}$ vector state from the
synchronization manifold, and consider the $m \times N$ column
vectors $\mathbf{X}=({x}_{1},{x}_{2},\ldots,{x}_{N})^T$ and
$\delta\mathbf{X}=(\delta {x}_{1},\ldots, \delta {x}_{N})^T$.

Then, one has

\begin{eqnarray}
  \delta\dot{\mathbf{X}} & = &
  \left[{\Bbb I}_{N}\otimes {J}{f}({x}_{s}) -
  \sigma
   {\cal L}\otimes {J}{h}({x}_{s})\right]\delta\mathbf{X},
  \label{eq:var}
\end{eqnarray}

where $Jf$ and $Jh$ represent the Jacobian operators of the system
dynamics and coupling. Following \cite{Pe:Ca}, Eq. (\ref{eq:var})
can be block-diagonalized in  the form:
\begin{equation}
\frac{d \eta_i}{dt}= [Jf(x_s)-\sigma \lambda_j Jh(x_s)] \eta_i,
\quad j=1,2,...,N, \label{blocks}
\end{equation}

where  $\{\lambda_j\}=\{\lambda_j^r+i \lambda_j^i\}$ is the set of
eigenvalues of the coupling matrix $\mathcal{L}$ (supposed
diagonalizable), ordered in such a way that $\lambda_1^r \leq
\lambda_2^r ... \leq \lambda_N^r$. Notice that $\lambda_1$ is
identically equal to $0$, due to the zero row-sum condition, and the
corresponding eigenvector is the unitary vector pointing in a
direction parallel to the synchronization manifold. Therefore, the
modes of the system along this direction are not relevant, since one
has only to care for whether the system will synchronize or not,
independent of where this happens in phase space. Thus, the
stability of the synchronization manifold should be checked with
respect to perturbations lying in phase space along directions
orthogonal to it (those associated to the eigenvalues
$\lambda_2,...,\lambda_N$).

In order to study the stability of the synchronization manifold as
a function of the eigenvalues of the matrix $\mathcal{L}$, one
introduces the following parametric equation in $\alpha$:
\begin{equation}
\frac{d \eta}{dt}= [Jf(x_s)-\alpha Jh(x_s)] \eta \label{parametric},
\end{equation}

where $\alpha$ takes the place of $\sigma \lambda_j$ in eq.
(\ref{blocks}). Note that $Jf$ and $Jh$ %represent the Jacobianoperators of the system's dynamics and coupling? Therefore, they
do not depend on the network topology. Moreover, for simplicity, one
can rewrite (\ref{parametric}) as:

\begin{equation}
\frac{d \eta}{dt}= K(\alpha)  \eta  \label{kernel},%\quad i=1,2,...N
\end{equation}
where $K(\alpha)$ is the evolution kernel, as a function of
$\alpha$. The MSF associates to each value of the parameter
$\alpha$ in the complex plane, the maximum (conditional) Lyapunov
exponent of the system calculated along trajectories of
(\ref{kernel}).

Now, in most cases (i.e. for most of the forms of the functions $f$
and $h$), the MSF is negative in a bounded region of the complex
plane (termed as the stability region $\mathcal{S}$). The immediate
consequence of this is that the network synchronizability is related
to the distribution of the eigenvalues of the coupling matrix
$\mathcal{L}$. Specifically, the stability of the synchronization
manifold requires as a necessary condition that all the $\sigma
\lambda_j$, $j=2,...,N$ belong to $\mathcal{S}$.

As proposed in \cite{Bocc2}, a sufficient condition to
verify the above requirement is checking that the rectangle of %defined on
the complex plane, $\mathcal{R}$: $[\sigma \lambda_2^r,\sigma
\lambda_N^r] \times [\sigma \min_j \lambda_j^i,\sigma \max_j
\lambda_j^i]$ is fully contained in $\mathcal{S}$ (observe also that
the spectrum of $\mathcal{L}$ is symmetric with respect to the real
axis and thus $M=\max_j \lambda_j^i=-\min_j \lambda_j^i$). Moreover
this condition can be simplified with respect to different forms of
the functions $f$ and $h$ into the general requirement that both the
eigenratio
defined as $R=\lambda_N^r/\lambda_2^r$ %($\lambda_i^r$ being the realpart of the eigenvalue $\lambda_i$)
and $M$, the maximum imaginary part of the spectrum of
$\mathcal{L}$, are as small as possible. In so doing, different
network topologies become directly comparable in terms of their
synchronizability, by simply looking at the indices $R$ and $M$.

Moreover, we wish to emphasize that there is no need for introducing
an ordering between the two parameters $R$ and $M$, defined above;
specifically, when they are both reduced, this leads to the general
result that the range of values of $\sigma$, say $I_{\sigma}$, for
which $\mathcal{R}$ is contained in $\mathcal{S}$ (and therefore the
network synchronizability) increases. Note that this result is
independent from the specific forms of $f$ and $h$, i.e. of the
particular MSF considered.
%
%the condition needed in order to increase the network
%synchronizability with respect to various forms of $f$ and $h$ (that
%is of the MSF) is that them both are made as small as possible.}

%, ensures if the spectrum $\sigma \lambda_i$, $i=2,...,N$ (which is
%symmetrical with respect to the real axis) is contained in the Note
%that it is possible the smaller is the area of the complex plane,
%which contains strictly
It is worth noting that the block-diagonalization in (\ref{blocks}) %this approach (as presented above) is properly %rigorously
can be properly applied only to the case of symmetric and asymmetric
diagonalizable matrices (in the particular case of a real spectrum,
as e.g. when $\mathcal{L}$ is symmetric, we obtain the following
simple relationship: the lower is $R$, the larger is the interval
$I_{\sigma}$). Moreover, as explained in \cite{Ni:Mo06,Lu:Ch04}, the
same definition of network synchronizability (in terms of both $R$
and $M$) can be extended also to the case of non-diagonalizable
matrices (see e.g. \cite{Mo:Zh:Ku04,paradox,Bocc1,Bocc2}), when the
condition is satisfied that the network embeds at least an oriented
spanning tree. The case of generic directed networks has been
studied in \cite{Wu04}.

%The above approach has been recently used
%\cite{enhancing,paradox,Bocc1,Bocc2} also in the case of asymmetric
%matrices, for which the condition that the modes associated with
%$\sigma \lambda_i$ ($i=2,...,N$) are orthogonal to the
%synchronization manifold is not in general satisfied, and for which
%the set of eigenvalues $\{\lambda_j\}$ is contained in the complex
%plane. In all those cases, it has been shown that, even though MSF
%criteria cannot be rigorously applied, they in fact still provide a
%very good qualitative description of the stability properties of the
%synchronization manifold.

Thus in what follows we shall seek to characterize how the
structural properties of complex networks, in terms of both the
forms of the adjacency matrix $\mathcal{A}$ and the weights matrix
$\mathcal{W}$, may affect (and eventually improve or hinder) their
synchronizability.

The first contribution in this sense was provided in \cite{Ni:Mo},
where unweighed scale free networks were compared in terms of their
synchronizability as varying the degree distribution exponent
$\gamma$.

Note that in the case of unweighed topologies, $\mathcal{W}$ is
assumed to be the unitary matrix and thus $\mathcal{L}$ can be
simply expressed as $\mathcal{L}_{ij}=-\mathcal{A}_{ij}$ for $i \neq
j$ and $\mathcal{L}_{ii}=\sum_{j} \mathcal{A}_{ij}$, $i=1,2,...,N$.

%$D-A$, where $D=\{D_{ij}\}$ is a diagonal matrix such that
%$D_{ii}=k_i$, $i=1,2,...,N$, where $k_i$ denotes the in-degree of
%node $i$.

It can be shown that as the exponent $\gamma$ is increased, the
network topology becomes more homogeneous and the average distance
within pairs of connected nodes along the geodesic increases. For
instance, real-world networks are characterized by low values of
$\gamma$ in that they are affected by high heterogeneity in the
degree distribution and low distances between pairs of connected
sites (the so-called \emph{small-world} effect).

Thus one would expect such networks to be characterized by high
synchronizability, in order also to justify their onset in nature as
being the result of a self-organized process. Instead, the analysis
in \cite{Ni:Mo} showed that these are typically characterized by
lower synchronizability than their homogeneous counterparts, giving
rise to an unexpected phenomenon, which is termed in the literature
\cite{paradox} as the \emph{paradox of heterogeneity}.

This first result was overturned in \cite{Mo:Zh:Ku04} where an
appropriate choice of the weights was considered over the network
links of the form:
\begin{equation}
\mathcal{W}_{ij}={k_i}^{-\beta}, \quad i \neq j, \label{beta}
\end{equation}
where $k_i$ is the degree of node $i$, i.e. $k_i = \sum_j
{\mathcal{A}_{ij}}$ and $\beta$ is a variable parameter.

 Now, it is worth noting that in the real
world, weights associated to the network links depend on an huge
amount of variables; furthermore they are typically also variable in
time. Thus, trying to estimate the effects of such a complex
behavior, may be a difficult task. On the other hand, the approach
proposed in \cite{Mo:Zh:Ku04} is very convenient in that it is aimed
at evaluating the effects of variable weights, depending essentially
on some network topological features. Namely, by varying $\beta$, it
is possible to tune the strength of the coupling among the network
nodes according to their degree: $\beta=0$ corresponds to the case
of unweighed topologies, positive (negative) values of $\beta$
indicate that the strengths of the couplings acting on each vertex
decreases (increases) with its degree. In \cite{Mo:Zh:Ku04} it was
shown that the network synchronizability is optimized at $\beta=1$,
whatever the form of the degree distribution. Specifically, it was
claimed that in this optimal regime, the eigenratio
$\lambda_N/\lambda_2$ is independent from the networks topological
properties, other than the average degree (note that for a coupling
of the form in (\ref{beta}), the parameter $M=0$ \cite{Mo:Zh:Ku04}).

The case of $\beta=1$ is of particular relevance also in another
aspect. In fact, in such a case, $\mathcal{L}_{ii}=1 \quad \forall
i$ and $\sum_{j \neq i} \mathcal{L}_{ij}=-1 \quad \forall i$.
Gerschgorin circle theorem \cite{Bocc2} states that the set of the
eigenvalues $\{\lambda_j= \lambda_j^r+ j \lambda_j^i \}$, belongs to
the union of the circles (${\mathcal{C}}_i$) centered on the main
diagonal values of $\mathcal{L}$ in the complex plane and with
radius equal to the sum of the absolute values of the other elements
in the corresponding rows, $(\{\lambda_i\}  \subset \bigcap_i {
\mathcal{C}_i [\mathcal{L}_{ii}, \sum_{j \neq i}
|\mathcal{L}_{ij}|]} )$. Thus in the case of $\beta=1$, all the
$\mathcal{C}_i$ coincide with the circle of radius $1$, centered at
1 on the real axis, inside which are constrained to lie all the
$\lambda_j$, $\quad j=1,2,...,N$ \cite{Bocc1}. Note that from this
condition, we can draw the two following ones: (i) $0=\lambda_1^r
\leq \lambda_2^r \leq ... \leq \lambda_{N}^r \leq 2$, (ii)
$\lambda_j^i \in [-1,1]$, $j=1,2,...,N$.

It is worth noticing here that for any given choice of the weights
over the network connections $\mathcal{W}_{ij}$, the same
requirements as above are satisfied by replacing (\ref{L}) with

\begin{eqnarray}
\mathcal{L}_{ij}=\frac{-\mathcal{A}_{ij}\mathcal{W}_{ij}}{\sum_k{\mathcal{A}_{ik}\mathcal{W}_{ik}}},
\quad i \neq j \nonumber\\
\mathcal{L}_{ij}=1, \quad  i=j. \label{normcoupling}
\end{eqnarray}

We wish to emphasize that this can be convenient in comparing
networks characterized by different topological features in terms of
their synchronizability. In fact, although the condition on the
minimization of $\lambda_N/\lambda_2$ corresponds to an increase of
the interval of the values of $\sigma$ for which the synchronization
manifold is stable, it does not give any information about the
values of $\sigma$ themselves. One possible consequence of this
could be e.g., that although a network is more synchronizable than
an other (in terms of their respective eigenratios), it experiences
synchronization only for higher values of the coupling $\sigma$.
Nonetheless this problem can be overcome by considering an
appropriate coupling as in (\ref{normcoupling}), which guarantees
that $\lambda_2$ is lower-bounded by $0$ and $\lambda_N$ is
upper-bounded by $2$ and thus ensures the spectra we are evaluating
are characterized by the same \emph{scale}.

By following this approach,  in \cite{Bocc1} it was shown that an
even better result in terms of network synchronizability may be
achieved by taking weights proportional to a specific topological
quantity known as \emph{load} \cite{korea} or \emph{betweenness
centrality} \cite{New:Bet}. At the same time in \cite{Bocc2}, a
particularly suitable way was proposed of choosing the weights over
the network connections, according to the presence of a given
ordering among the network nodes. Specifically the network nodes
were ordered according to a given scalar property, say $p$, in such
a way, i.e. that $i \leq j$ if $p(i) \leq p(j)$. Then the following
general way of choosing the weights was proposed:

\begin{eqnarray}
\mathcal{W}_{ij}=\frac{Q_{ij}}{\sum_{k} \mathcal{A}_{ik} Q_{ik}},
\end{eqnarray}
where $Q_{ij}= \frac{1-q}{2}$ ($Q_{ij}= \frac{1+q}{2}$) if $i<j$
($i>j$), with $-1<q<1$ being a parameter governing the coupling
asymmetry within the network. Note that, by varying $q$, it is
possible to study the effects of variable asymmetric weights over
the network edges. Specifically, in what follows we will consider a
very natural way of choosing the weights based on a degree ordering
(also termed as age ordering in \cite{Bocc2}), i.e. such that
$p(i)=k_i$. Namely, $q=0$ corresponds to the case of symmetric
coupling, which is equivalent to the choice of $\beta=1$ in eq.
(\ref{beta}). The limit of $q \rightarrow 1$ ($q \rightarrow -1$)
corresponds to the case of a direct network in which only the
connections from high-degree to low-degree vertices (from low-degree
to high-degree) are present. Under this condition, it was shown in
\cite{Bocc2}, that in the case of scale-free networks,
synchronizability can be improved for decreasing values of $q$, i.e.
when the strength of the coupling is dominating in the direction
from the high-degree to the low-degree vertices.

Sofar, we have considered the effects of varying the weights over
the network connections  on the network synchronizability (in terms
of the matrix $\mathcal{W}$), while the network topology  (in terms
of the matrix $\mathcal{A}$) has been considered as fixed.
Nonetheless, it was shown in \cite{Ni:Mo,IJBC_NDES}, that even in
the case of unweighed topologies, synchronizability may be strongly
affected by varying some particular network topological properties.

The most important one is, as already stated before, the degree
distribution, which affects in a non-negligible way almost all the
dynamical processes taking place over networks \cite{report}. On the
other hand, many important properties have been discovered to
characterize in more detail the network structure, other than the
degree distribution. These are mostly due to particular forms of
correlation or mixing among the network vertices \cite{New03Mix}.

One form of mixing is the correlation among pairs of linked nodes
according to some properties at the nodes. A very simple case is
degree correlation \cite{New02Ass}, in which vertices choose their
neighbors according to their respective degrees. Non-trivial forms
of degree correlation have been experimentally detected in many
real-world networks, with social networks being typically
characterized by assortative mixing (which is the case when vertices
are more likely to connect to other vertices with approximately the
same degree) and technological and biological networks, by
disassortative mixing (that takes place when connections are more
frequent between vertices of different degrees). In \cite{New02Ass}
this property has been conveniently measured by means of a single
normalized index, the Pearson statistic $r$ defined as follows:
\begin{equation}
\label{eq:r} r={1
\over{\sigma^2_q}}{\sum_{k,k'}kk'(e_{kk'}-q_kq_{k'})},
\end{equation}
where $q_k$ is the probability that a randomly chosen edge is
connected to a node having degree $k$; $\sigma_q$ is the standard
deviation of the distribution $q_k$ and $e_{kk'}$ represents the
probability that two vertices at the endpoints of a generic edge
have degrees $k$ and $k'$ respectively. Positive values of $r$
indicate assortative mixing, while negative values characterize
disassortative networks.

In \cite{IJBC_NDES} a strategy was used, similar to the one
presented in \cite{New02Ass}, which allows to vary the degree
correlation (in terms of variable values of $r$), while keeping
fixed the degree distribution and it was shown that disassortative
mixing, i.e. the tendency of high-degree nodes to establish
connections with low-degree ones (and viceversa), is indeed a
desirable network property in terms of its synchronizability. The
same effect was checked in \cite{SUB} to be persistent under a
normalized form of the weighing as that proposed in
\cite{Mo:Zh:Ku04}, with $\beta=1$.

Here, we shall seek to characterize how the combination of the
effects of variable degree correlation in terms of the structural
configuration of the network, and variable asymmetry over the
network weights may indeed affect the network synchronizability.

In order to construct networks, characterized by a given degree
distribution (in terms of the exponent $\gamma$), we consider the
network building model proposed in \cite{Mo:Re95}, which is
commonly termed as the \emph{configuration model}. Specifically
this consists in the following steps:

\begin{enumerate}
  \item Start from a given form of the degree distribution $P(k)$.
  \item Assign to each of $N$ vertices a \emph{target degree} $k>k_{min}$ drawn from the distribution
  $P(k)$.
  \item Randomly create connections between pairs of vertices, avoiding multiple and self-connections, until all vertices have reached
  their \emph{target degree}.
\end{enumerate}

Then in order to reproduce a desired level of degree correlation,
following \cite{New03Mix}, we propose exchanges of links between
pairs of connected nodes, until the desired value of the observable
$r$ is reached.

\begin{figure}
\centering
\includegraphics[width=7cm]{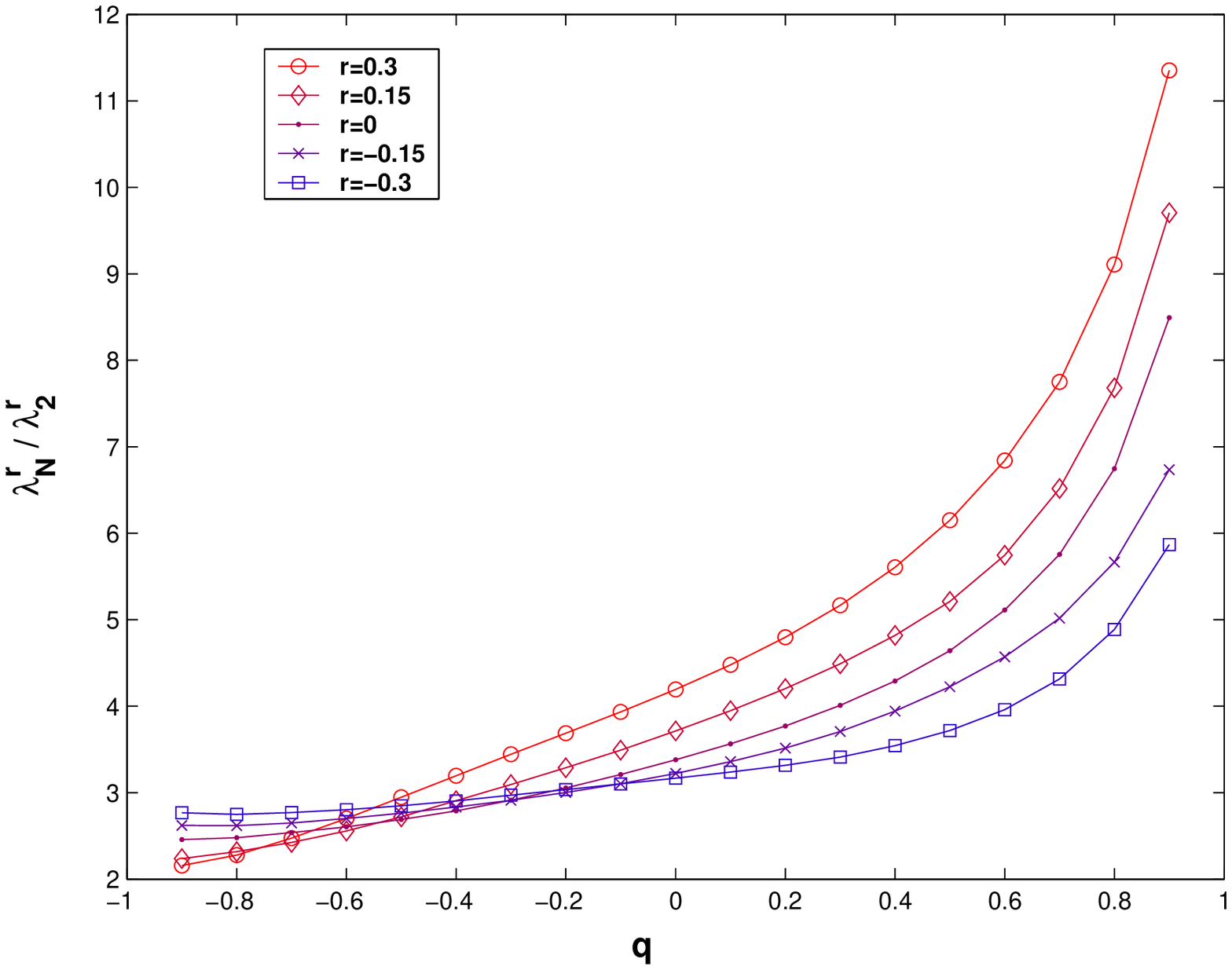}
\includegraphics[width=7cm]{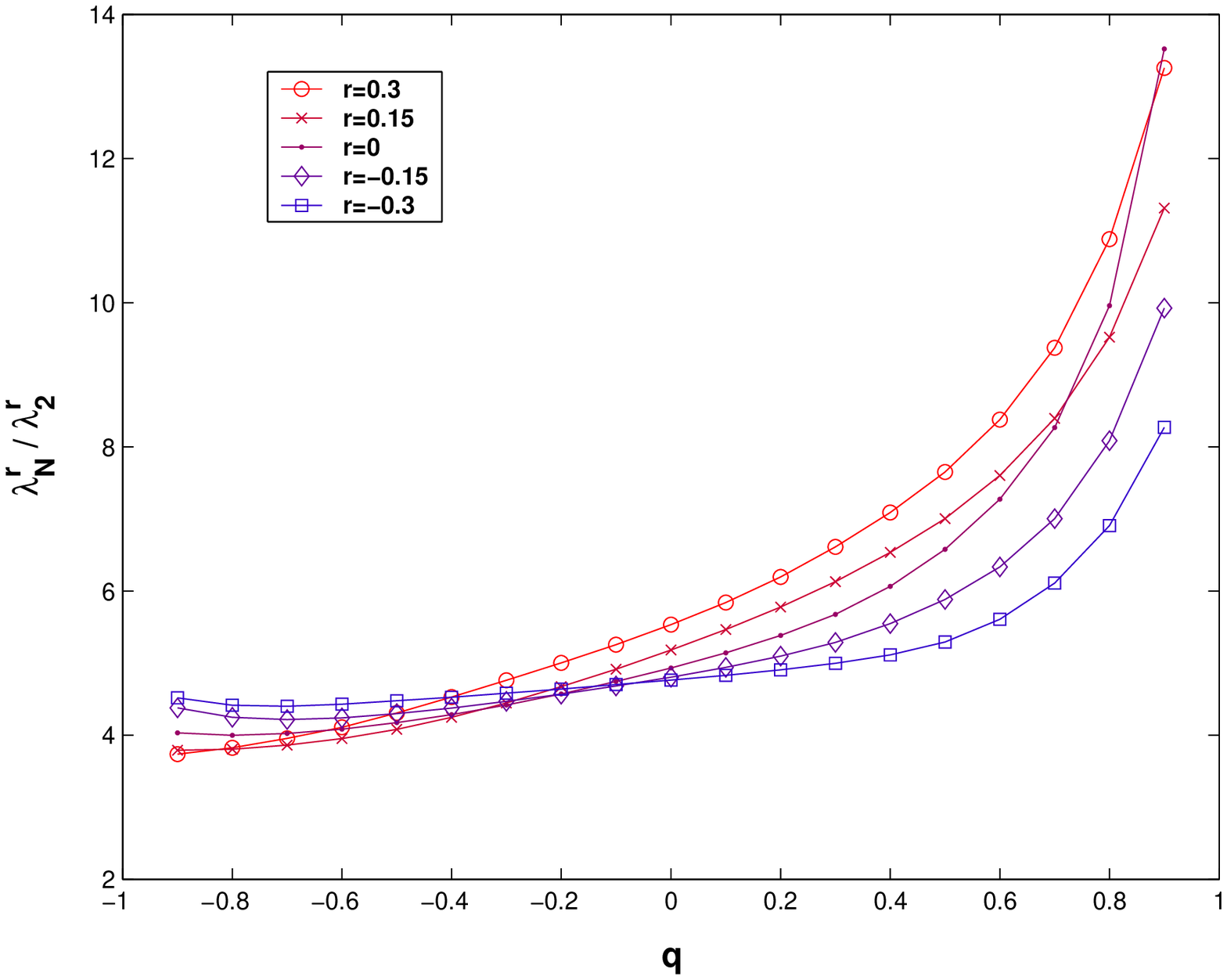}
\caption{Eigenratio $\lambda_N/\lambda_2$ as function of q for scale
free networks characterized by variable degree correlation:
$r=[-0.3,-0.15,0,0.15,0.3]$. Left panel: $\gamma=2$, $N=1000$. Right
panel: $\gamma=3$, $N=1000$. \label{figura1}}
\end{figure}

\begin{figure}
\centering
\includegraphics[width=7cm]{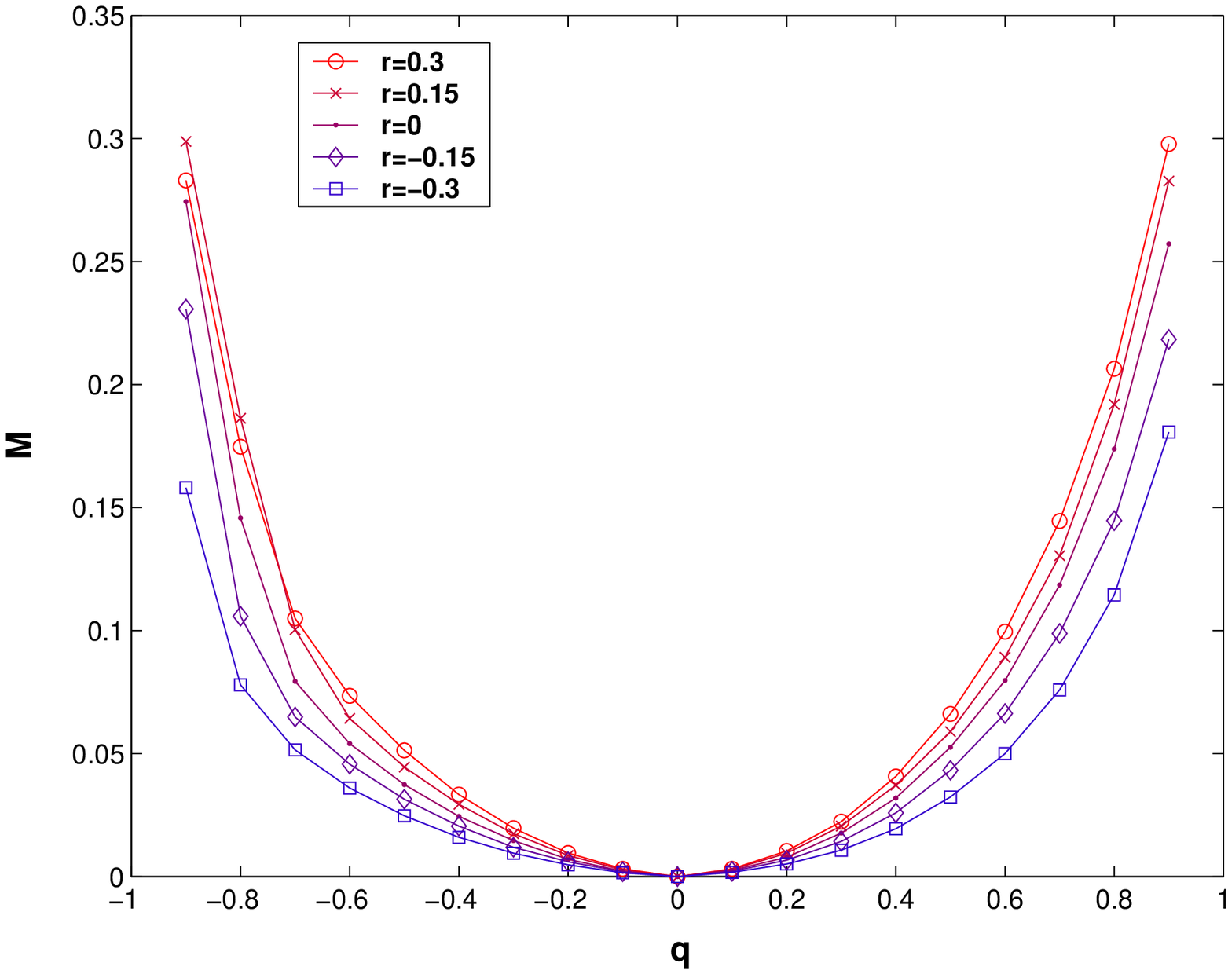}
\includegraphics[width=7cm]{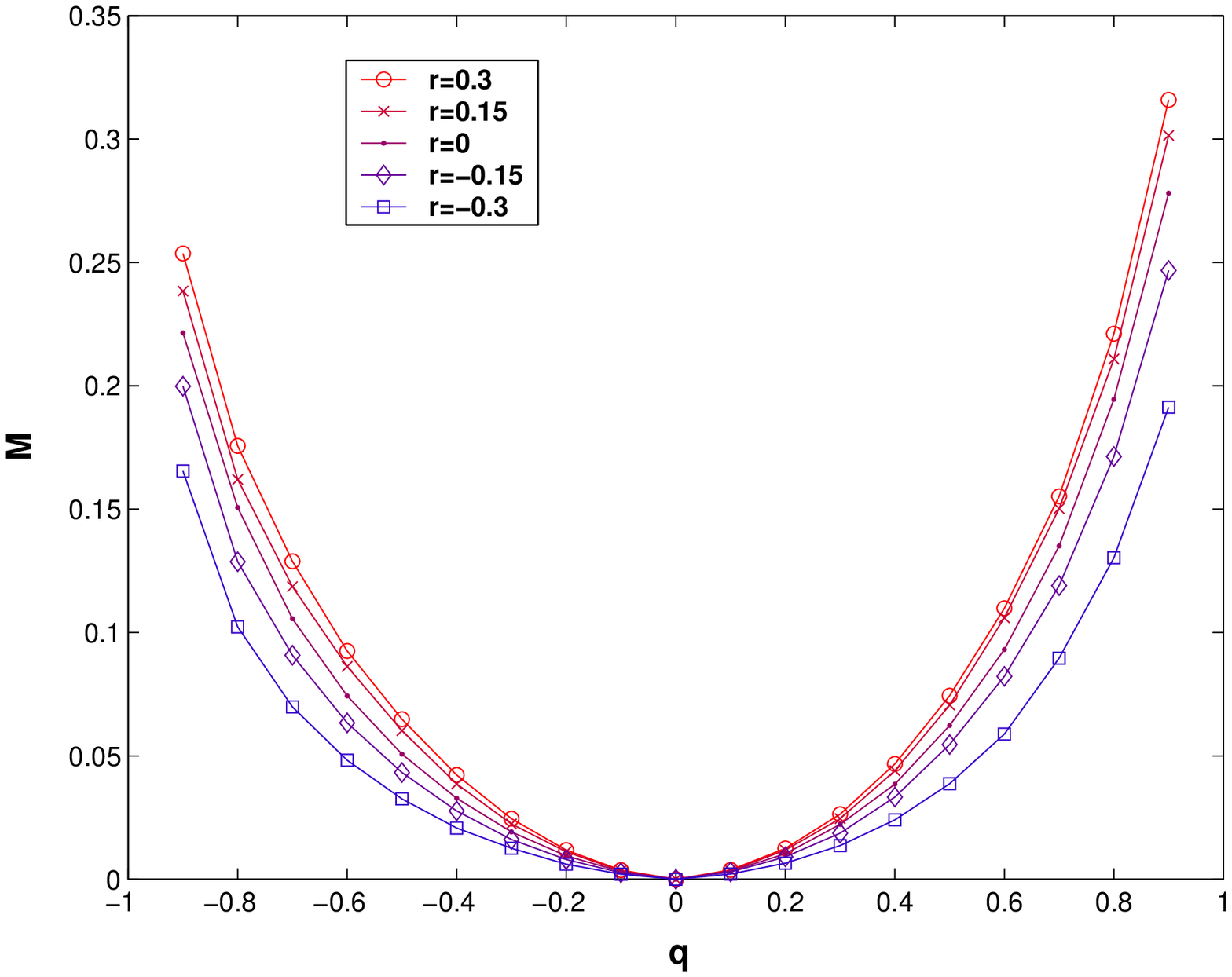}
\caption{Maximum imaginary part of the spectrum of $\mathcal{L}$, M,
as function of q for scale free networks characterized by variable
degree correlation: $r=[-0.3,-0.15,0,0.15,0.3]$. Left panel:
$\gamma=2$, $N=1000$. Right panel: $\gamma=3$, $N=1000$.
\label{figura2}}
\end{figure}

In what follows we have evaluated the effects on the network
synchronizability of both variable values of $r$ and $q$. The main
results are shown in Figs. \ref{figura1} and \ref{figura2}, for
scale free networks with degree distribution exponent
$\gamma=\{2,3\}$.

The effects on the eigenratio $R$ have been reported in Fig.
\ref{figura1}. Note that, as already shown in \cite{Bocc2} for
uncorrelated networks, $R$ is an increasing function of $q$ for
all the values of $r$ considered. This depends on the advantage,
in terms of the network synchronizability, of having asymmetric
interactions directed from high-degree nodes to low-degree ones.
Moreover this is particularly evident for low values of $q$ in the
case of assortative networks. Thus the combination realized when
$q \rightarrow -1$ and the network topology is %assumed to be
strongly assortative (high $r$), represents the optimum
configuration for the minimization of $R$.

This is also consistent with the qualitative claim presented in
\cite{Bocc2} that the optimal network configuration is obtained,
when both a dominant interaction from high-degree nodes to
low-degree ones and a structure of interconnected hubs are present.

\begin{figure}
\centering
\includegraphics[width=7cm]{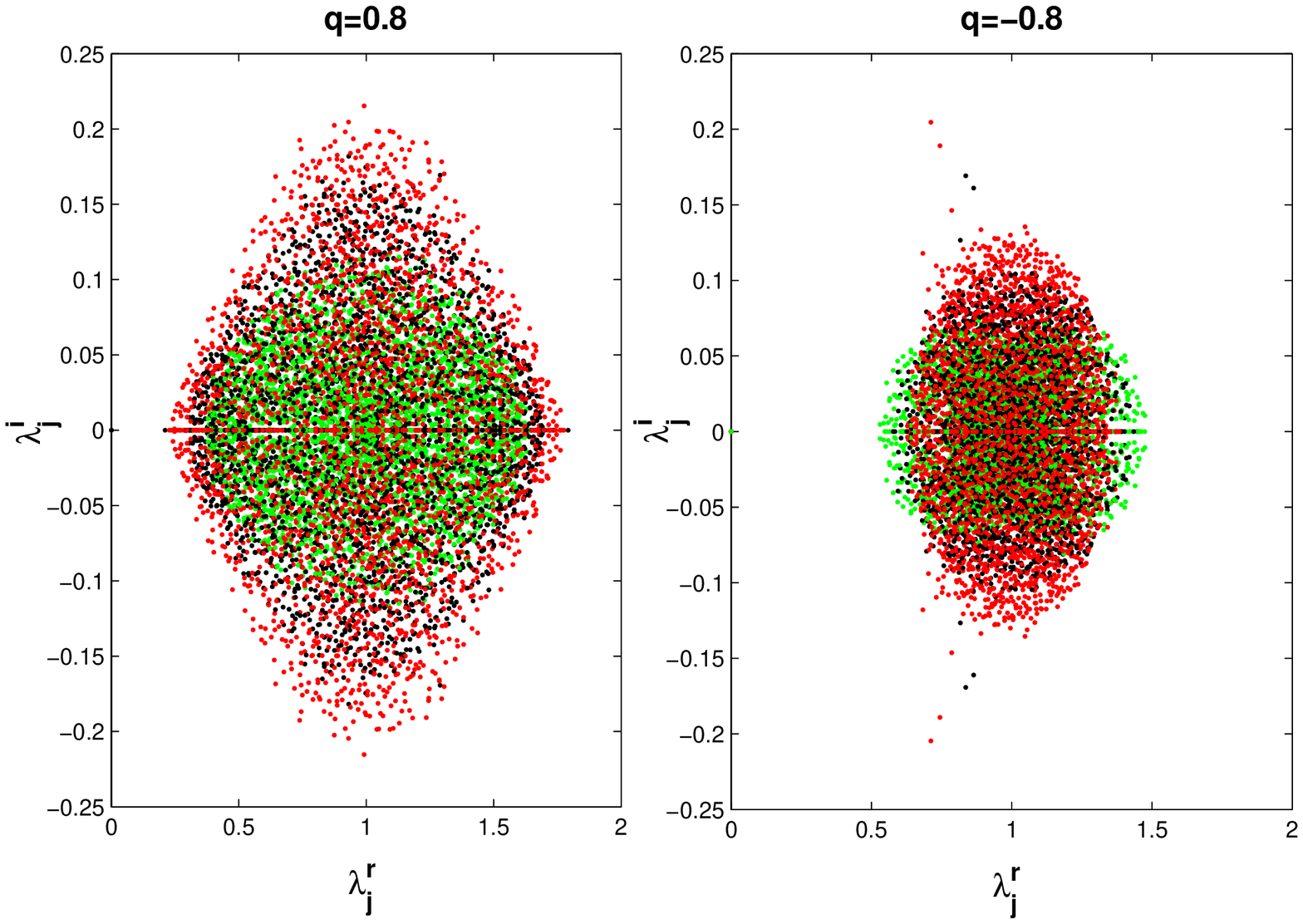}
\includegraphics[width=7cm]{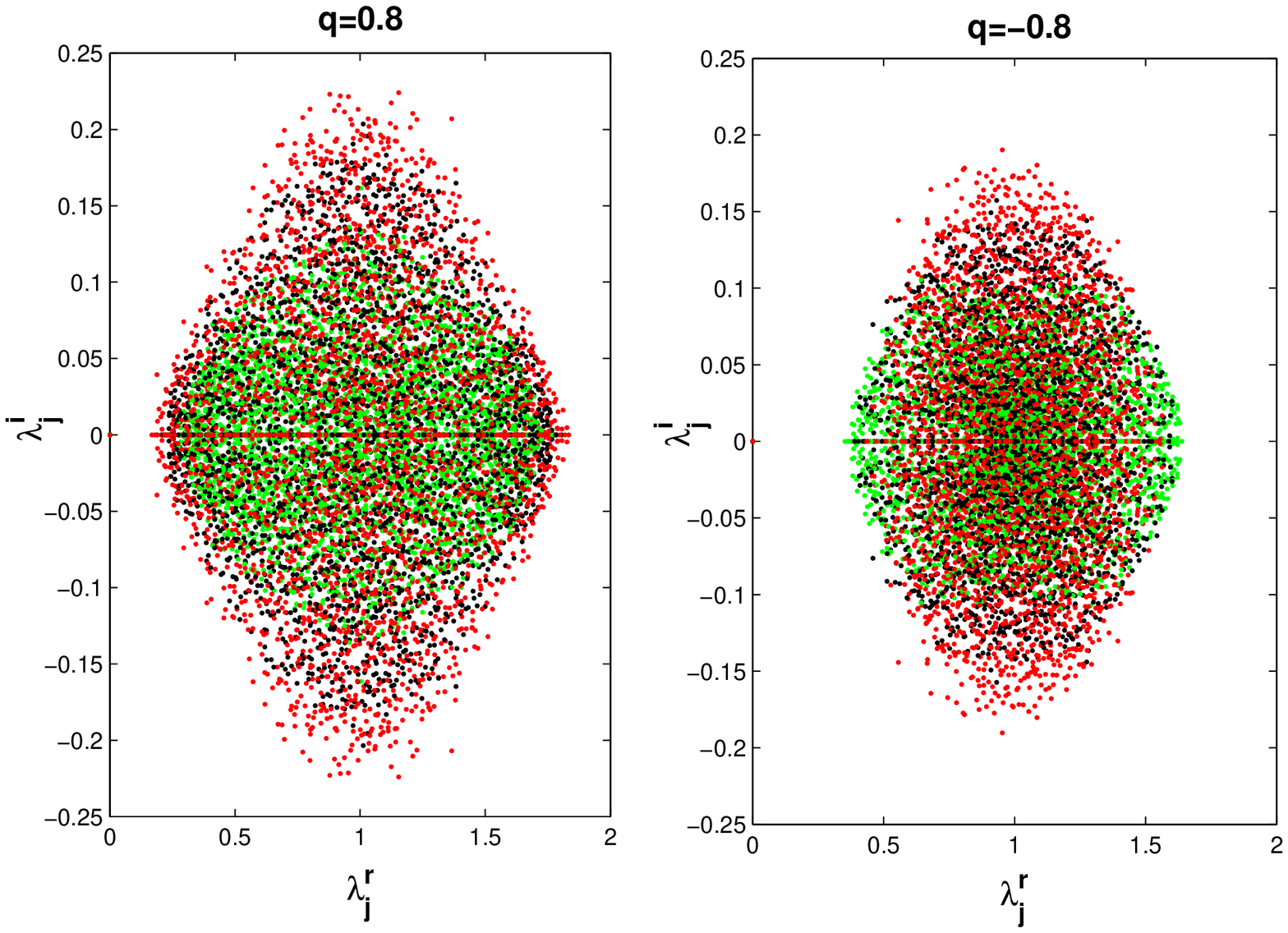}
\begin{picture}(0,0)(0,0)
\put(-382,120){\small ${(a)}$} \put(-289,120){\small ${(b)}$}
\put(-180,120){\small ${(c)}$} \put(-80,120){\small ${(d)}$}
\end{picture}
\caption{Spectrum of networks characterized by different degree
correlation properties and degree of asymmetry. Each picture
represents $\lambda_j^i$ vs $\lambda_j^r$, $j=1,2,...,N$. Green
points are used for the spectrum of networks characterized by
negative degree correlation ($r=-0.3$), black points are used for
the spectrum of networks characterized by absence of correlation
($r=0$), red points for the spectrum of networks characterized by
positive degree correlation ($r=+0.3$): (a) $\gamma=2$, $q=0.8$; (b)
$\gamma=2$, $q=-0.8$; (c) $\gamma=3$, $q=0.8$; (d) $\gamma=3$,
$q=-0.8$. \label{figura3} }
\end{figure}

On the other hand, when $q \approx 0$, we observe a completely
different picture with disassortative networks being characterized
by better synchronizability properties (as already shown in
\cite{SUB}). The same behavior is also confirmed in the case of
positive values of $q$, where the already better performance of
disassortative networks is further enhanced through asymmetric
coupling.

Thus the onset of two separate regimes emerges as varying $q$: the
first one, in the case where there is a dominant interaction of
the high-degree nodes on the low-degree ones, with assortative
mixing representing
 a desirable property in terms of the network synchronizability;
 the second one, with symmetric coupling or directed from low-degree nodes to high-degree ones, where
 disassortative mixing represents the best choice in order to enhance the network synchronizability.

The behavior of the maximum imaginary part of the spectrum, $M$,
as varying $q$, is show in Fig. \ref{figura2}. As expected, this
is characterized by a minimum at $q=0$ (for which the coupling is
symmetric) and increases as increasing the asymmetry over the
network links. Note that for all values of $q \neq 0$, assortative
(positively correlated) networks are characterized by higher
values of $M$ when compared to their uncorrelated and
disassortative counterparts.

A more exhaustive picture is provided in Fig. \ref{figura3}.
Specifically here %in Fig. \ref{figura3},
two configurations
characterized by strong asymmetry ($q=0.8$ and $q=-0.8$) are
compared for scale free networks with exponents $\gamma=\{2,3\}$, in
terms of their spectra: $\lambda_j^r$ vs $\lambda_j^i$, for
$j=1,2,..,N$. Networks characterized by different degree correlation
properties are depicted with different colors: red for assortative
mixing, green for disassortative, and black for no-correlation.
Remind that the condition for increasing the network
synchronizability is that the spectrum $\{\lambda_j\}$ for
$j=2,..,N$ is contained in a region of the complex plane as small as
possible.
%\emph{as compact as possible}.

This condition is clearly achieved for negative values of $q$
(compare respectively  Figs. \ref{figura3}(a) with \ref{figura3}(b)
and Figs. \ref{figura3}(c) with \ref{figura3}(d)). Note that in the
case of $q=0.8$, the advantage of introducing negative degree
correlation is undoubted, as both $R$ and $M$ are reduced and the
green area is clearly smaller than the red one in Figs.
\ref{figura3}(a) and \ref{figura3}(c). On the contrary, in the
optimal regime where $q=-0.8$, the reduction of the range of the
real values of the spectrum $\lambda_j^r$, $j=2,3,...,N$ is
counterbalanced by an increase in the range of the imaginary parts
$\lambda_j^i$, $j=2,3,...,N$, so that a more intricate phenomenology
emerges.

 Specifically, in such a case it becomes necessary to look at
the specific form of the MSF under evaluation, in order to state the
benefit of introducing degree correlation among the network
connections. This should be taken into account, also with respect to
variable forms of the MSF, when different choices of the functions
$f$ and $h$ are taken into consideration.

\begin{figure}
\centering
\includegraphics[width=10cm]{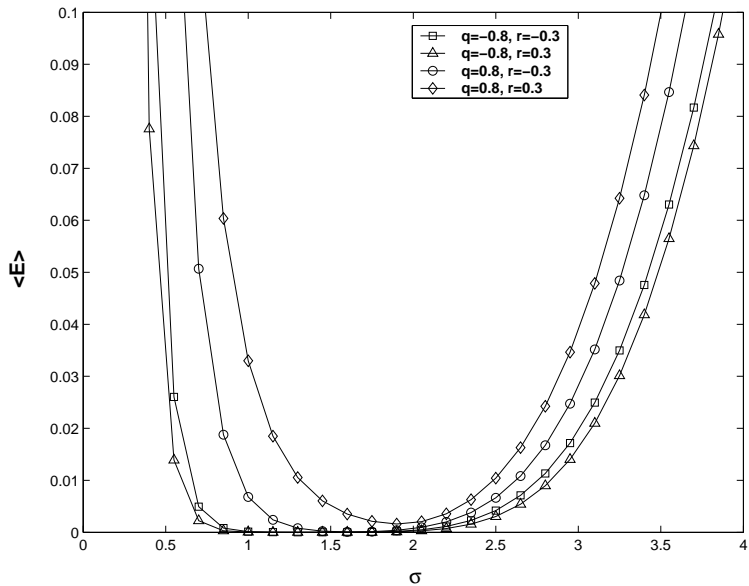}
\caption{Synchronization error $ \langle E\rangle$ %(see text fordefinition)
as a function of the coupling strength $\sigma$ for a
scale-free network of 1,000 R\"ossler oscillators
(\emph{configuration model}, $\gamma=2$). Evolution equations and
parameters are specified in the text. Each point corresponds to an
ensemble average over a set of 10 independent realizations. The
asymmetry and correlation parameters are $q=0.8, r=0.3$ (diamonds),
$q=0.8, r=-0.3$ (circles), $q=-0.8, r=0.3$ (triangles) and $q=-0.8,
r=-0.3$ (squares). The numerical simulations support the qualitative
scenario of the MSF: for positive (negative) values of $q$,
synchronization is enhanced in disassortatively (assortatively)
correlated graphs. \label{figura5}}
\end{figure}

In order to confirm this qualitative description with an example, in
the following we numerically simulate a network of 1,000 coupled
chaotic  R\"ossler oscillators. The dynamics of the system is ruled
by Eq. \ref{eq:net}, with $m=3$,  ${x} = (x_1,x_2,x_3)$,
${f}(x)=[-x_2 - x_3, \ x_1 + 0.165 x_2, \ 0.2 + x_3(x_1-10)]$ and
${h}(x)=[x_1, 0, 0]$.

The appearance of the synchronous state can here be monitored by
looking at the vanishing of the time average (over a window $T$) of
the synchronization error $ \langle E\rangle =
\frac{1}{T(N-1)}\sum_{j>1}
  \int_t^{t+T} ||{x}_j-{x}_1||dt'$. In the present case, we
  adopt as vector norm $||{x}||=|x_1|+|x_2|+|x_3|$.

Fig. \ref{figura5} reports $\langle E\rangle $ {\it vs}. $\sigma$
for different networks topologies and degree correlations. It is
easy to observe that the numerical simulations fully support the
qualitative scenario of the MSF. Indeed, by comparing the
synchronization effects in the cases $q=0.8, r=0.3$ (diamonds),
$q=0.8, r=-0.3$ (circles), $q=-0.8, r=0.3$ (triangles) and
$q=-0.8, r=-0.3$ (squares), one immediately realizes that for
positive (negative) values of $q$, synchronization is enhanced in
disassortatively (assortatively) correlated graphs.

In conclusion, in this paper we have regarded the network topology
as the set of \emph{active connections} present in the network, that
is pairs $\{ij\}$, $i,j=1...N$, such that $\mathcal{A}_{ij}=1$. We
analyzed the effects of varying the weights over the network
connections (by making the parameter $q$ change) in order to
investigate different levels of asymmetry. Moreover, we studied how
variable degree correlation, while keeping the degree distribution
unchanged, can favor or hinder the overall synchronization. While
negative degree correlation was found to be a convenient network
structural property in terms of synchronizability in the case of
symmetric coupling ($q=0$) and of positive values of $q$
%asymmetrical interactions from low-degree to high-degree vertices(i.e. $q>0$)
, a more intricate phenomenology was observed in the optimal
situation, i.e. for low values of $q$. Specifically, in such a case
an interesting balance between the width of the range of the real
and imaginary parts of the eigenvalues has been observed.

We wish to emphasize that this issue could be of relevance in those
applications where, because of different circumstances,
synchronization may be or not be a desirable effect.

\end{document}